# Mobility Management Framework

Péter Fülöp[a,*], Benedek Kovács[b,*], Sándor Imre[c,*],

[a,c]Department of Telecommunications, Budapest University of Technology H-1521 Pf. 91,Hungary,
[b] Department of Mathematical Analysis, Budapest University of Technology H-1111, 1 Egry József Hungary

**Abstract**

*This paper investigates mobility management strategies from the point of view of their need of signalling and processing resources on the backbone network and load on the air interface. A method is proposed to model the serving network and mobile node mobility in order to be able to compare the different types of mobility management algorithms. To obtain a good description of the network we calculate descriptive parameters from given topologies. Most mobility approaches derived from existing protocols are analyzed and their performances are numerically compared in various network and mobility scenarios. We developed a mobility management framework that is able to give general designing guidelines for the next generation mobility managements on given network, technology and mobility properties. With our model an operator can design the network and tune the parameters to obtain the optimal implementation of course revising existing systems is also possible. We present a vertical handover decision method as a special application of our model framework.*

## 1 Introduction

Information mobility has became one of the most common services in the modern world with the widespread of the portable phones and other mobile equipments. The wireless multimedia and other services has many requirements and the resources in the serving network are often expensive and limited.

*Email address:* {fulopp,imre}@hit.bme.hu,bence@mcl.hu (Péter Fülöp[a,*], Benedek Kovács[b,*], Sándor Imre[c,*]).

We investigate mobility as an abstract problem regardless of actual technical solutions or serving network. In the first mobility protocol designs, the main scope was to create a well-functioning mobility. For example, the Global System for Mobile Communication (GSM) network uses a Cellular approach to save bandwidth on the air interface but does not really focus on the problem of signalling load on the wired serving network. In the Mobile IP (MIP) [10] structure the IP mobility is in the main scope. There are many enhancements of MIP to optimize the original protocol and introduces for example hierarchy, location tracking to obtain a solution that is more cost efficient in a way. There are various other technologies, too. Host Identity Protocol (HIP) [5] is drastically different from MIP although they have similar centralized mobility approach but are implemented on a different network layer. The Wireless Local Area Networks (WLAN) are constructed similarly to the original Local Area Networks (LAN) and provide mobility only within the radio interface and uses Dynamic Host Configuration Protocol (DHCP). Future protocols might use different media and technological background to provide mobility.

The advantage in our work is that we do not focus on a selected technology not even on a given network generation but discuss mobility in general within the modern computer and telecommunication networking technologies. To describe the environment we focus on the bottlenecks such as the air interface, bandwidth on the core network and processing load on the network nodes and as one of the most significant properties: complexity involving scalability.

We compare selected mobility approaches and show how the network properties affect the usability of each. The aim is to find the suitable one for different scenarios and to give guidelines how to construct the network for a protocol or a protocol to the network. To generalize the model we have to uniform the notations as well. The cooperating mobility management elements in our approach will be the Mobile Nodes (MNs) attaching to Mobility Access Points (MAPs) as a subset of the network of Mobility Agents (MAs).

This paper is structured as follows. The model for the network and the mobile node with its mobility parameters are introduced in Section 3. This is followed with the definitions of the cost functions for existing approaches in Section 4. One can see figures of the numerical results in Section 5 while the conclusion is derived in Section 6.

## 2 Mobility Management

In this paper the mobility management is discussed generally regardless of the particular technology used. We try to grab the most significant properties of the mobility that is worth to discuss within the scope of the modern mobility protocols.

## 2.1 The Mobility Management System

We define *Mobility Management System – MMS* as an application running on network nodes that helps to locate the mobile equipment towards its unique identifier (the IP address for example).

- The Mobile Nodes (MN) are the mobile equipments who want to communicate to any other mobile or fixed partner.
- There are Mobility Access Points (MAP) as the only entities that directly connects the Mobile Equipments. (Note: mobility does not necessarily imply radio communication. It means only that the Mobile Node changes its Mobility Access Points and when it is attached to one, connection between them can be established.)
- The Mobility Agents (MA) are network entities running the mobility management application.
- There is a core network that provides communication between the Mobility Access Points and has a structure that can be described with a graph. Vertices are either Mobility Access Points or Mobility Agents other serving nodes who are not part of the mobility management application and the edges can be various links (even radio links) for the data communication between the vertices.

With this definition one can see that most of the functional entities of the current mobility protocols and others under development can be generally described. (Details and proof of the above statement is given in Section 4.)

We simplify the problem of mobility management to a protocol that finds the correct, marked Mobility Access Point where a given Mobile Node is attached. To create this model we need some practical assumptions:

- A Mobility Access Point is always a Mobility Agent.
- All the nodes presented above are logical entities i.e. Mobility Access Points can mean a set of physical access points.
- A mobile equipment can communicate with multiple Access Points at the same time but one connection is necessary and enough to maintain the correct communication. The mobile can also attach and detach from any Mobility Access Points. At this point we assume that the mobile node is administrated only at one agent. This means that the problem of finding the mobile node is the same as finding the correct access point. (It is not difficult to enhance the discussion to the case where the MN can be find at multiple MAPs but it is out of the scope of the paper.)
- The nodes in the core network communicate and find each other using a

given protocol or method (for example via IP routing). For this reason this part of the mobility protocols is not discussed.

The above definition and assumptions suit our aim that is to investigate the properties of various management strategy approaches, since the number of messages sent and the number of tasks should be completed can be calculated. With assigning appropriate cost parameters to each message or task one will be able to model exact mobility management solutions and can analyze them. We are going to give an example in Section 5.

## 2.2 Mobility Strategy Approaches

Here we present a classification of Mobility Management Systems from the signalling strategy point of view. We distinguish between five main approaches that are the centralized, hierarchical, wireless and wired tracking and cellular ones.

These main approaches we will investigate are derived from existing protocols for example the centralized-like learnt from the MIP [10] or Session Initiation Protocol (SIP) or Host Identity Protocol (HIP) systems. However, the Cellular IP (CIP) [2] as an adaptation of Global System for Mobile Communication (GSM) is used mainly in the micro mobility layer, it is possible to further extend it hierarchically through the whole network and we refer to this class as the cellular approaches. Also we will discuss hierarchical-like solutions such as (Hierarchical Mobile IP - HMIP [3], Telecommunication Enhanced Mobile IP - TeleMIP [13]) as obvious extensions of the centralized-like approaches to resolve scalability problems. Tracking-like solutions both wired (HAWAII [12]) and wireless tracking (Tree Location Area - TrLA [11], Location Tracking - LTRACK [8]) are based upon the idea that the former point of attachment could forward the call or the packet in the direction of the new MAP where the mobile moved to.

There can be some other special approaches not fitting exactly to any of these but its model is going to be easy to construct using the following examples. We give an example for such a process in Section 4.4. It is also common that the mixture of applications is used on different mobility layers. One can easily model, analyze and tune his mobility protocol or network, using our framework. With our investigations we believe that designing guidelines for new generation network mobility protocols can be given.

# 3 Network Graph and Node Mobility Parameters

In this Section, we introduce how we will model the network structure on which the Mobility Management Systems work. To derive the main parameters we will have to model the behavior of the Mobile Nodes as well. There will be general and algorithm specific parameters derived and also we will describe how we handle some protocol specific, extra problems like *looping* (how likely that a mobile returns to a former MAP) and the frequency of *location area changes* (how likely that a MN moves within a given subset of MAPs).

Secondly, we present the cost dimensions include to our model. These cost parameters related to the "signalling on the links" as a bandwidth and interworking equipment usage (or even QoS and Service Cost ratio), the "processing in the nodes" which are taken into account only on the nodes running the mobility protocol, the "access cost" or "air interface usage" containing the cost of accessing the fixed network (MAP-MN attachment) or explicitly for example the battery consumption of the MN.

## 3.1 Modelling the Network

As the first step we go through existing comparison works because in many papers, the network is modelled in order to emphasize the properties of a single protocol compared to another one. This approach is not flexible since new protocols cannot be imported to the comparison and also the little modifications in the protocols are difficult to follow but a model of this type sometimes essential for a protocol. Let us see the two main approaches of network models.

One approach to describe the network is to give global parameters like a general average distance between nodes. It is true with this approach, any kind of network could be described since the parameters can mostly be derived although the method to derive them is often not presented in the works. However, as we will see, introducing these parameters is not enough to compare most of the protocols because they cannot emphasize the benefits of each. (See [9] as an example.)

Other works model the network with given network structure so each protocol is examined in the environment it was designed into. Clearly, it is often essential to make appropriate examination but it makes difficult to extend the discussion. For example, when a GSM cell structure is used, no vertical handovers are taken into account: another mobility protocol might have a different structure of covering the same geographical region when the graph, describing the network might not even be able to be drawn on a plane that will be the case in our model. (See [11] as an example.)

Summing up the requirements we introduce a method to derive global parameters from any kind of network to get the benefits of the first approach and we show a method how the protocol specific structure parameters can be derived. This generalizes the discussion while keeping some important specific characteristics.

## 3.2 Deriving Parameters of a Given Network

Let us have a given network topology with a given MN behavior. The network is modelled with a graph and so are the possible movements of the mobile nodes. Thus the initial model is a weighted adjacency matrix for the network and a handover frequency (intensity) matrix for the Mobile Nodes movement.

### 3.2.1 MN behavior and position, - Handover Frequency Matrix

Let us assume that the aggregated behavior of the Mobile Nodes can be modelled with a finite state continuous Markov chain (the handover or call arrival rate than is a Poisson process with various intensity parameters as in many works, e.g. [4]). The chain is given with a rate matrix $B_Q = [b_{ij}]$. In this matrix, all the possible (in practice: the practically possible) MA-s are listed where the Mobility application runs. (These MAs can also denote single access points, bigger networks or the Home Agent if desired.) The number of MAs is $n$ and so the matrix will be an $n \times n$ matrix where each element $b_{ij}$ denotes how frequent the movement of the mobile is from $\text{MAP}_i \to \text{MAP}_j$. If an MA is not a MAP then there are 0 values in its row and column (i.e. we threat it the same way that the MN cannot or never attaches to it).

From the rate matrix the transition matrix $B_\Pi$ can be determined easily. We assume that the matrix $B_\Pi$, without the non-MAP nodes, is practically irreducible and aperiodic that implies that the chain is stable and there exists a stationary distribution. This will be denoted by a density vector $\underline{b}$. In this vector, the $i$th element denotes the probability of the MN being located under the $i$th MAP. This is the same probability as the relative number of handovers from and to the $i$th MAP. (For MA nodes that does not support access point functionality, there is an element in the vector with 0 value.)

### 3.2.2 Network, - Network Matrix

Let us have the corresponding network graph given with its weighted adjacency matrix: $A$. This matrix should include all the nodes in the network where the mobility application runs (all the MAs again) so has the same $n \times n$ size as matrix $B_Q$ and $B_\Pi$. The weights are relative values thus $w_{ij} = 1, w_{jl} = 2$ means that the relative cost (or any kind of measure) of any mobility management signalling from $j \rightarrow i$ is the double of the cost from $i \rightarrow j$. Several interpretations of the weights are possible for example relative delay or jitter or the number of routers on the path, etc.

With the Floyd algorithm the optimal distances between the nodes can be calculated (even with weighted or directed edges as well). The distance between nodes will be the sum of weights on the shortest (cheapest) path from one to the other. Let this result matrix be given by $A_d$. In the $i$th row of the matrix, the distances from MA$_i$s are listed. Let the distances from the HA, - a special MA, - be given with the vector $\underline{a}$.

We will parameter $w$ to denote the average of the weights in the network. It can be calculated by summing up the elements of $A$ and dividing it with $n^2$.

### 3.2.3 Determining m

Parameter $m$ will denote the average depth level, that is the average sum on weighted edges on the shortest (cheapest) path from the MN to the HA. Clearly, the average number of vertices among the path is $m+1$ if $w_{ij} = 1 \forall i, j$.

We will use matrix $A_d$ and vector $\underline{a}$ to calculate this parameter. Both have to be normalized with the average weight of edges in the network ($w$). Now $mw$ can be calculated with determining the weighted average of the distances where the weights are the probabilities that the node is under a given MAP.

$$m = \frac{\underline{a} * \underline{b}}{w}, \tag{1}$$

where $*$ stand for the scalar product. One can see that the nodes which are not MAPs have a 0 multiplier and do not count in the average distance as expected.

These parameters $m, mw$ we learnt show the real average number of edges and distances along the shortest path from the HA. If there is a call or delivery request, we suppose that it is routed on this optimal path towards the current MA of the MN. In this sense these are the smallest values of our parameters. (After calculating $m$ one can further manipulate it for example multiply it with the probability of choosing the optimal path for each protocol if needed, etc.)

### 3.2.4 Parameter $g_T$

We will have another parameter like $m$ that is the average distance between two nodes who handles the MNs handovers. They might be connected, but they can also be quite far from each other logically due to different technologies especially in the case of vertical handovers. So as we see this parameter has to denote the weighted average value of the length between every two neighboring MAs where the mobile can attach. Than it is calculated as follows:

$$g_T = \frac{\underline{b} * tr(A_d \cdot B_\Pi)}{w}. \tag{2}$$

We denote it with $g_T$ since this parameter will have the most effect on the Tracking-like management solutions as we will see.

### 3.2.5 Parameter $g_H$

This parameter denotes how far is the nearest hierarchical junction to register in average if we consider the optimal covering tree of the network with the HA in the root. The junction node is the nearest common node of the paths from HA to the old and the new FA of the MN. In Figure 1 MN moves to position 2, in this case the nearest hierarchical node is MAP B, and $g_H$ represents the distance between MAP A and MAP B (In most cases, the optimal tree structure is not possible to achieve since the different service providers will not mesh their networks: approximate values can be used instead.)

Our hierarchical structure will be built up using two main parameters. First is the number of nodes: $n$. The other parameter is the average number of neighboring MAs that can be accessed via a wire from a given node: $\delta$. It should be also weighted with the probability density of the MN.

$$\delta = \frac{(\underline{b} \cdot sign(B_\Pi)) \cdot \underline{1}}{n}. \tag{3}$$

From this, our parameter $g_H$ will be computed using the approach to compute $g$, derived in [8].

### 3.2.6 Parameter $g_C$

This parameter will denote the average distance of MAPs from the main MA of a Location Area in the Cellular-like approaches. Together with this parameter we can compute the number of MAPs in a cell ($n_C$) and the average probability of making handovers out from the cell ($p_C$). These will be introduced later.

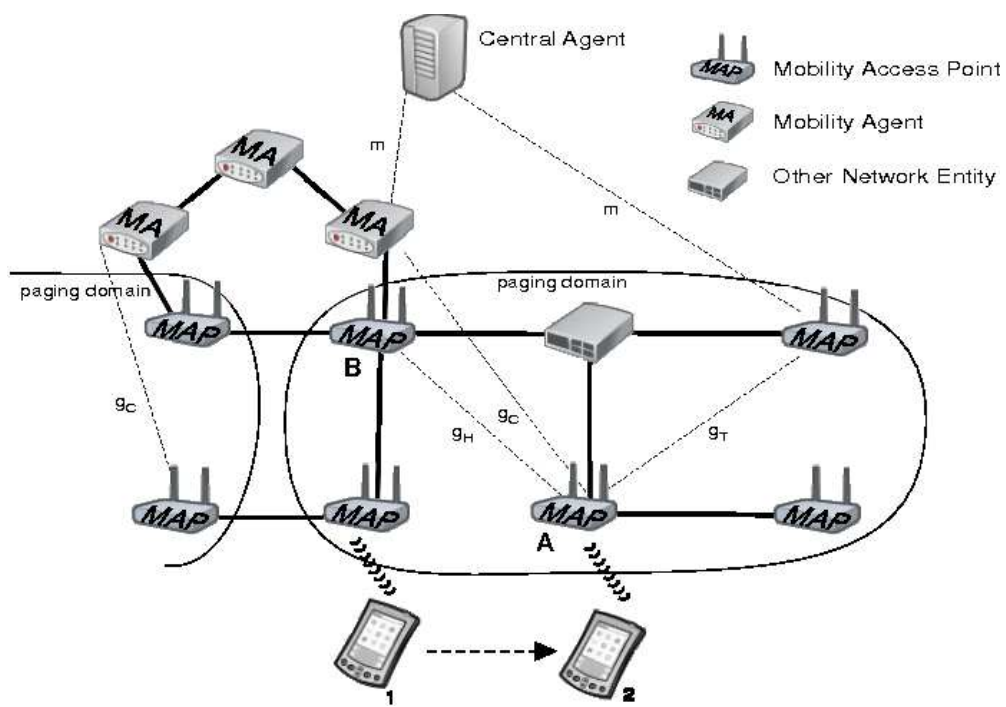

Fig. 1. Presentation of the network parameters

It is an NP full problem to calculate the optimal cell structure, but there are algorithms approaching it very good in some sense. For example [16] solves this problem under additional constraints and limits the maximal paging cost. In our numerical simulation we have run the algorithms developed and published in [14] and obtained $g_C$ with them. However, concerning the NP fullness two important notes have to be mentioned. When talking about an existing network, these parameters can be calculated easily. If we work with small network patterns, the optimal cell structure can be selected from the not too many options.

### 3.3 Modelling the Mobile Node

As we have seen, matrix $B_Q$ describes the movement behavior of the MN, handover-wise. If one sums up the $i$th row in this matrix it gets a rate how frequent the MN moves from the $i$th MA (MAP) with a Poisson-process. Let $\lambda$ denote the average parameter of the Poisson-process (at each MAP) and so denote the rate of handovers for a general MN anywhere in the network.

The other parameter that can be introduced in a similar manner is the rate of receiving a call: $\mu$. This parameter can also be time- or location-dependent. We take its average value as we did in the case of $\lambda$ and we assume it is constant in the examined very small time interval just like we did in the case of matrix $B_Q$ and through the whole modelling.

Using the achievements in [8], let us introduce $\rho$ as the "mobility ratio" meaning the probability that the MN changes its FA before a call arrives. The good thing in our notation is that various movement modelling can be embedded into it with varying $B_Q$.

### 3.4 Loop Removal Effect and Moving Within a Subset of MAPs

We include the movement directions of the mobile node because many protocols try to exploit the advantage of returning to the a previous point of attachment or staying within a range of nodes. The first gives significant advantage of tracking protocols while the second has the effect on the performance on every protocol especially on the cellular approaches except the centralized-like ones. The *loop removal* is discussed within the optimization of the Tracking-like and Cellular-like protocols. We use the achievement presented in our previous work on LTRACK mobility management [8]. Returning to a subset of node is still open for investigations but party incorporated to the problem of Paging Area decisions where we used one of our colleagues' method [14].

## 3.5 Definitions of Cost Constants

The three main classes of cost types and the corresponding cost constants that can describe the technology level will be introduced here. Why do we need distinguished cost types?

The main reason is that the same kind of protocol can be implemented on various network layers that influences its signalling and processing needs of each protocol message. Another reason to introduce such network topology and mobility strategy independent cost constants is that the underlying networking equipments might have very different characteristics so it might be important to test the behavior of the Mobility Management System on different serving networks to find the most suitable one. For these reasons, one will always have to tune these parameters according to the very implementation used. For our numerical example values see Section 5 where we also show that modifying the ratio of some (or some set of) parameters (for example the registration and packet forwarding cost) strongly affect the performance of a mobility management system.

### 3.5.1 Link Related Constants

First class of the cost constants are the "link related constants" (see. Table 1). They are introduced since one of the most important properties of a network is the bottleneck of uplink and downlink bandwidth especially when the service has to satisfy Quality of Service (QoS) requirements as well.

As it was described, our network model does not necessarily show the real network topology. Each edge in the graph denotes the link from one MA to another. There might be several routers and subnetworks among the path. The parameters introduced here gives one unit signalling cost in each direction.

### 3.5.2 Node Related Constants

The "node related constants" model the cost of resources in the MA nodes (the vertices of our graph). For example the cost of creating a packet might be different for different protocols (number of headers, if there is tunnelling, etc.) so these constants should be adopted to the examined protocol but also might be different using equipments of different vendors too.

Table 1
The cost contants

| Signalling constants | $c_u$ | The unit cost of one update on a link. |
|---|---|---|
| | $c_d$ | The unit cost of one delivery on a link. |
| Processing constants | $c_r$ | Registration cost, the cost of the process in the MAP when a MN node wants to attach. |
| | $c_f$ | Forwarding cost at a MA. |
| | $c_m$ | This is the constant cost of modifying some node related records in a MA. |
| | $c_{ec}$ | The cost of building up a message. |
| | $c_{rc}$ | The cost of recapsulating or rebuliding a message. |
| | $c_{dc}$ | The cost of decapsulate or open the message at an endpoint. In many cases $c_{rc} = c_{dc} + c_{ec}$. |
| Air if. constants | $c_{au}$ | The cost of uplink message between the MN and the MAP. |
| | $c_{ad}$ | The cost of downlink messaging between the MN and the MAP. |

### 3.5.3 Mobile Equipment Connection Related Constants

The "access interface and connection related constants" will denote of the unit cost of using the "access interface". This access interface can denote radio access or the cost when one connects his laptop to an ethernet cable and there is load on the network from the laptop (MN) to the MAP (server of the network). We will collect the costs of this group of tasks in two parameters: upload and download.

It is easy to see that the cost might be different for upload, download especially if download is a simple broadcasted paging (mainly MAP resources needed) while upload is a registration message that needs more resource (battery) at the MN side. As we mentioned in Section 3.1 the layer 2 handoff costs, for example the movement detection, Access Point (AP) searching, and AP reassociation can be taken into account here.

### 3.5.4 Example for cost diversification

Here we try to underpin the above classification theoretically. We want to give an example but of course many kinds of setups are possible. When a MN moves (leaves its old MAP and connects to a new one) different events happens in layer 2 and layer 3. The costs of these events, the layer 2 handoff, the agent discovery and the actual registration are incorporated to interface and air interface cost in our paper. The first one depends on the access technology and can be accomplished in different ways. For example in the case of IEEE 802.11b WLAN it means the AP changes, and can be split into three parts: movement detection, AP searching, and AP reassociation. In order to handle this difference of layer 2 handoff of access technology we can take into account this cost using an average layer 2 handoff cost. Agent discovery is another process in layer 3, during this phase, the Mobility Access Point advertises their services on the network by using a Discovery Protocol for example ICMP Router Discovery Protocol (IRDP) in Mobil IP. The Mobile Node listens to these advertisements to determine where it is connected to, or if it is connected to its home network or foreign network. The Discovery Protocol advertisements can carry the types of services MAP will provide such as reverse tunneling and Generic Routing Encapsulation (GRE) and the allowed registration lifetime or roaming period for visiting Mobile Nodes. Rather than waiting for agent advertisements, a Mobile Node can send out an agent solicitation. This solicitation forces any agents on the link to immediately send an agent advertisement. When the Mobile Node hears an Agent advertisement and detects that it has moved outside of its last network then registers its current location during registration process.

## 4 Modelling the Existing Approaches

In this Section, the five main classes of mobility management protocols introduced in Section 2 are shortly described and modelled with their signalling-, processing-, and air interface cost functions.

The cost function as the model for the mobility approach can be any kind of utility function depending on the purpose of use. It is constructed as a calculation of expected utility as a function of network topology descriptors and cost constants with respect to mobility i.e. the conditioned probability of changing the attachment point with $\lambda(MAP_{ij}, t)$ intensity on not being paged with $\mu(t)$ intensity generally:

$$C = E[\int_{-\infty}^{\infty} f_h(n,c) dP(\lambda(MAP_{ij}, t)) \qquad (4)$$
$$+ \int_{-\infty}^{\infty} f_p(n,c) dP(\mu(t))],$$

where $\lambda, \mu$ are the intensities of a Poisson process and $n \in \{\text{"network parameters"}\}, c \in \{\text{"technology constants"}\}$. Functions $f_h, f_p$ are unique for each mobility management strategy and mostly are a linear function of the network parameters and the technology constants. They tell us the cost of one handover or one paging for each protocol respectively. Since the integral is taken with respect to a Poisson process it will be a simple sum and with introducing the handover

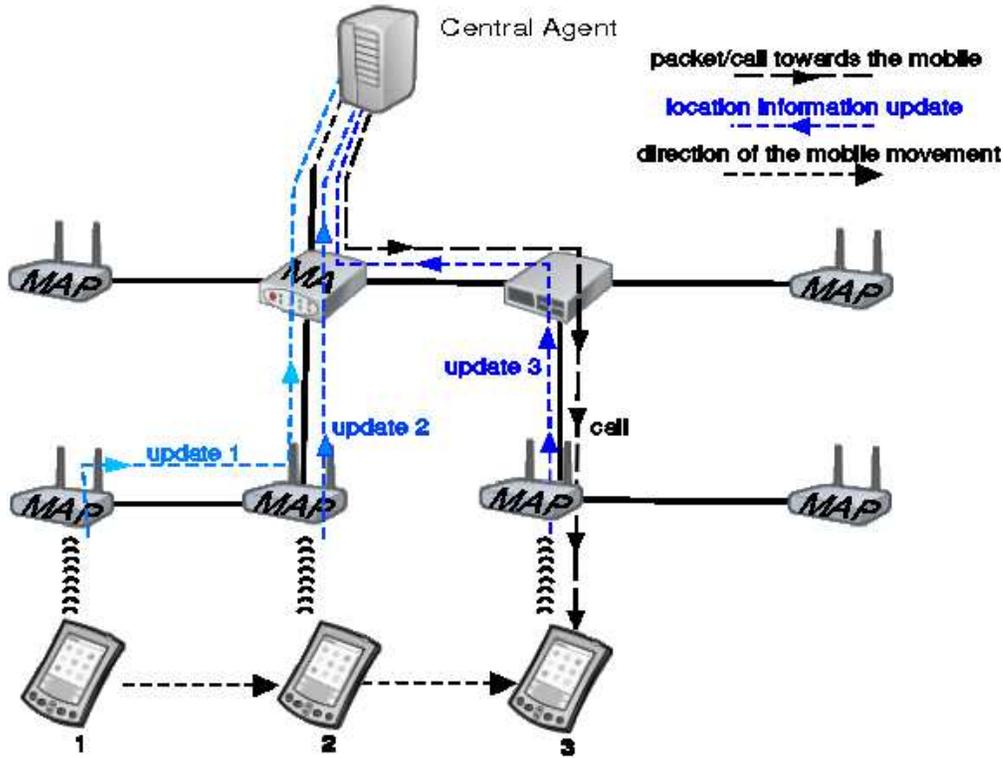

Fig. 2. Basic operation of Centralized Approach

process conditioned on no paging the above equation simplifies to.

$$C = \rho f_h + \rho f_p, \qquad (5)$$

where $\rho = \frac{\lambda}{\lambda+\mu}$ ($\lambda$ is the average in the network.) Further explanation and description about the cost function construction can be found in [8].

### 4.1 Centralized Approaches

This group of managements contains various protocols like Mobile IP [10]. The common in these management structures is that the MN always sends location update messages in case of handover to a central or a central group of management nodes, that maintains a database and has up-to-date information about the exact location of the MNs. (Figure 2.)

The Central Agent (or Central Agents) always has to know the exact location of the Mobile Node in order to inform the correspondent nodes, or to forward the requests to MN using tunnel or source routing. The followed method by Central Agent depends on the mobility solutions. For example in Mobil IP protocol the Home Agent as a Central Agent, forwards the packets using tunnel, but in SIP based mobility the Registrar server answers the IP address of MN to the SIP Proxy, which controls the communication.

Table 2
The cost functions

| Approach | Cost | Approach - Cost type function |
|---|---|---|
| Cent | Sig. | $\rho m c_u + (1-\rho)(mc_d)$ |
| Cent | Proc. | $\rho(c_r + (m-1)c_f + c_m) +$ |
|  |  | $(1-\rho)(c_{ec} + (m-2)c_f + c_{dc})$ |
| Cent | Air | $\rho c_{au} + (1-\rho)c_{ad}$ |
| Hier | Sig. | $\rho g_H c_u + (1-\rho)(mc_d)$ |
| Hier | Proc. | $\rho(c_r + (g_H - 1)c_f + c_m) + (1-\rho)(c_{ec} +$ |
|  |  | $(m - g_H - 1)c_f + c_{rc} + (g_H - 1)c_f + c_{dc})$ |
| Hier | Air | $\rho c_{au} + (1-\rho)c_{ad}$ |
| WlessT. | Sig. | $\rho P_H g_H c_u +$ |
|  |  | $(1-\rho)(g_H c_d + M[h_r]g_T c_d + (1-P_0)g_H c_u)$ |
| WlessT | Proc | $\rho((1 - P_H)(c_r + c_m) + P_H(c_r + (g_H - 1)c_f +$ |
|  |  | $c_m)) + (1-\rho)(c_{ec} + (m-1)c_f + P_0 c_{dc} +$ |
|  |  | $(1-P_0)(M[h_r]((g_T - 1)c_f + c_{rc}) +$ |
|  |  | $c_{dc} + (g_H - 1)c_f + c_m))$ |
| WlessT | Air | $\rho((1-P_H)2c_{au} + P_H c_{au}) + (1-\rho)(c_{ad})$ |
| WiredT | Sig. | $\rho(g_T(1-P_H) + g_H P_H)c_u +$ |
|  |  | $(1-\rho)(mc_d + M[h_r]g_T c_d + (1-P_0)g_H c_u)$ |
| WiredT | Proc. | $\rho(c_r + (g_T - 1)c_f + c_m) +$ |
|  |  | $(1-\rho)(c_{ec} + (m-1)c_f + P_0 c_{dc} +$ |
|  |  | $(1-P_0)(M[h_r]((g_T - 1)c_f +$ |
|  |  | $c_{rc}) + c_{dc} + (g_H - 1)c_f + c_m))$ |
| WiredT | Air | $\rho c_{au} + (1-\rho)c_{ad}$ |
| Cell | Sig. | $\rho(1 - P_{cell})g_H c_u +$ |
|  |  | $(1-\rho)(((m - g_C) + (n_C g_C))c_d + g_C c_u)$ |
| Cell | Proc. | $\rho((1-P_{cell})(c_r + g_H c_f + c_m)) + (1-\rho)(c_{ec} +$ |
|  |  | $(m - g_C - 1)c_f + c_{rc} + (g_C - 1)n_C c_f + n_C c_{dc})$ |
| Cell | Air | $\rho((1-P_{cell})c_{au}) + (1-\rho)(n_C c_{ad} + c_{au})$ |
| HPage | Sig. | $\rho(1 - P_{cell})g_C c_u +$ |
|  |  | $(1-\rho)((m - g_C)n_d + (n_C g_C c_d) + g_C c_u)$ |
| HPage | Proc. | $\rho((1-P_{cell})c_r + g_C c_f + c_m) + (1-\rho)(c_{ec} +$ |
|  |  | $(m - g_C - 1)n_d c_f + c_{rc} + (g_C - 1)n_C c_f + c_{dc})$ |
| HPage | Air | $\rho((1-P_{cell})c_{au}) + (1-\rho)(n_C c_{ad} + c_{au})$ |
| Manet | Sig. | $\rho(1 - P_{cell})g_H c_u +$ |
|  |  | $(1-\rho)((m - g_C + 1) + (P_M n_C g_C c_d) + g_C c_u)$ |
| Manet | Proc. | $\rho((1-P_{cell})c_r + g_H c_f + c_m) + (1-\rho)(c_{ec} +$ |
|  |  | $(m - g_C)c_f + c_{rc} + P_M g_C n_C c_f + c_{dc})$ |
| Manet | Air | $\rho((1-P_{cell})(g_C - 1)c_{au}) +$ |
|  |  | $(1-\rho)(P_M n_C g_C c_{ad} + c_{au})$ |

As we have just seen on Table 2 the cost functions are obvious and simple. In the processing function there is only one encapsulation, one decapsulation and $m - 2$ forwarding cost, which do not need a big computative capacity. The application of such a protocol has a second main advantage along with its simplicity, namely these approaches can be built by installing a Central Agent in the network and by running an IP-level software module on the MN. There is no need to change any other entity in the network, therefore it is cheap and easily installable. But we have to take a relevant disadvantage into account, centralized mobility puts extraordinary high overload on the bearer network and uses non optimal routing, which are unacceptable. However, this solution is far from the optimal, still the most of the mobility implementations use the same kind of this centralized approach. Transport layer mobility, for example mSCTP (multi-homing Stream Control Transmission Protocol), HIP and Application layer mobility for example SIP, as it was mentioned above, belong to this centralized approach. (The mSCTP allows an association between two end points to span multiple IP addresses or network interface cards. It supports to keep alive the TCP sessions during IP address change. HIP separate the node identification and location information in the network layer. This solution affords the applications a permanent network layer. SIP provide the possibility of many Voice over IP services.)

Despite these advantages all of these protocols need a centralized management to accomplish the mobility, which is far from the efficiency it could have.

### 4.2 Hierarchical Solutions

Instead of the global management node regional management system can be used to reduce the signalling traffic by maintaining the location information locally. For this reason we can use the MAPs and MAs as local agents, that have database to store the actual IP addresses of MN. So we can consider this hierarchical network structure as a tree of MAP, MA and other network node with Central Agent in the root of the tree. The main idea is that the update information is sent only to the nearest MA on the network, if the MN moves within the subnetwork managed by this entity. Parameter $g_H$ denotes this distance as it is explained in Section 3.2.5.

Let us now take a look at the network tree of the hierarchical mobility. One path from the HA leads to the old MA of the MN and another one leads to the new one. It is enough to send the location update message to the nearest common router with the old path and the new path, as one can see on the Figure 3.

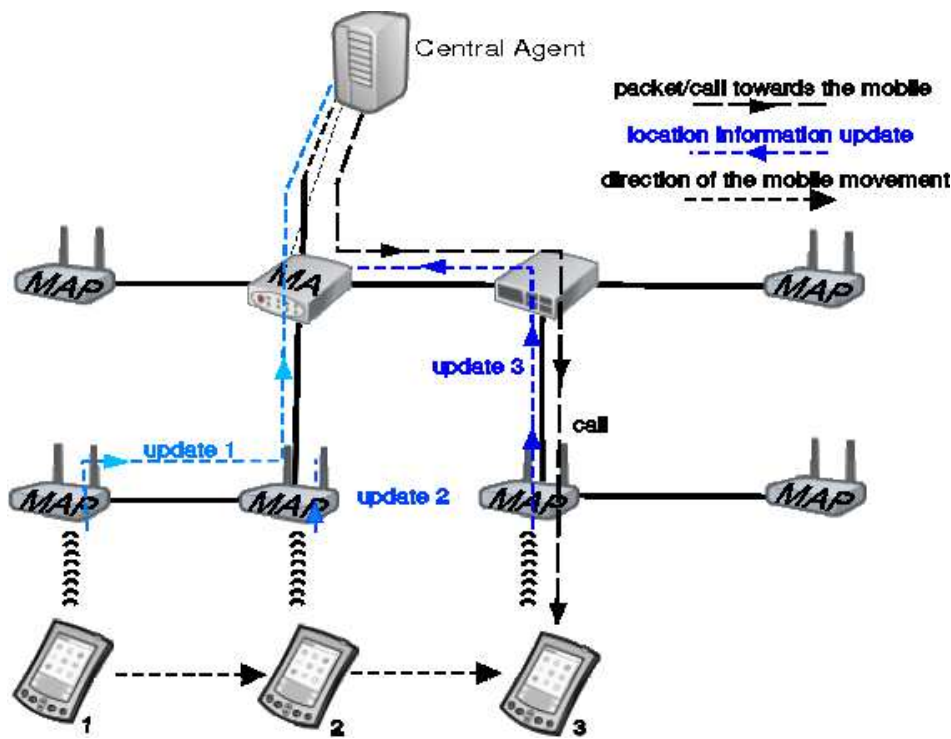

Fig. 3. Basic operation of Hierarchical approach

However, typically there are other nodes placed between the MAs. Now we take the best case for the signalling optimization and consider every node in the tree as a MA, or a set of MAs and MAPs. Within the subnetwork controlled by one MA a subnetwork IP is assigned to the node, and changes when the node changes its point of attachment in this level of the tree. In this approach Central Agent knows the IP address of the MA, under that in the network tree the MN is located, or even moves. The costs function changes compared to the centralized solution, because the location information is sent only to the nearest MA ( $g_H$ distant from the MN) .

The advantage of this method is the more optimal functionality, and less load on the bearer network. However, the change of some other entity is needed in the network, therefore the solution is more expensive. An example for such solution is the Hierarchical Mobile IP (HMIP) [3].

### 4.3 Tracking-like Solutions

In the tracking-like approaches each mobile node has an entry in a Central Agent like in other solutions. This CA stores the address where it received location update message from. It is the address of an MAP, and a next-hop towards the mobile node. The mobile node is either still connected to that MAP, or that MAP knows another next-hop MAP towards the mobile. Finally the mobile node can be found at the end of a chain of MAPs.

As it was introduced the main idea behind the tracking-like algorithm is that if the MN changes its point of attachment then it could be a good solution to send an update to the old MAP. The MAP that the mobile node moves away is called old MAP, the one it moves to is called new MAP. After this handover, called tracking handover the old MAP is able to forward the request towards the MN via the new MAP.

There is another kind of handover in tracking-like solutions, the normal handover. Normal handover occurs when mobile equipment updated its entry in the Central Agent by sending the address of the new MAP node to it.

The normal handover is similar to centralized solutions; it generates a lot of signalling traffic, but a tracking handover puts less, or no signaling to the network (see Section 4.3.2 and 4.3.1). But if an incoming packet arrives to the mobile node, we have to find it in a hop-by-hop manner, and send a location update message to the Central Agent, which are expensive. In tracking scheme a normal handover can be followed by some tracking handovers before another normal handover takes place. If the mobile node does not receive a packet between two normal handovers then less signalling is used, if a packet is received after some tracking handovers, more signalling is used compared to a centralized mobility scheme.

Thus the most important decision of tracking-like mobility is when to make a normal handover, and when to make a tracking handover. In the tracking-like approaches each mobile node has an entry in a Central Agent like in other solutions. This CA stores the address where it received location update message from. It is the address of an MAP, and is a next-hop towards the mobile node. The mobile node is either still connected to that MAP, or that MAP knows another next-hop MAP towards the mobile. Finally the mobile node can be found at the end of a chain of MAPs. One can read more about these protocols in works: [1], [12], [8].

We distinguish between two different tracking-like solutions based on the kind of tracking handover: wireless tracking and wired tracking. In case of tracking handover of wireless tracking the mobile sends the address of the new MAP node to the old MAP node over the air interface (for reference see LTRACK [8]) while in case of the wired tracing the information is sent over the wired network [12].

The optimal number of tracking handovers between two normal handovers has to be calculated in both cases. We use the achievements in [8] to determine these parameters and define the optimal cost function. The model and the effect of *loop-removal* is imported to our work as well.

### 4.3.1 Wireless Tracking

In case of tracking handover of wireless tracking the mobile sends the address of the new MAP node to the old MAP node over the air interface (Figure 4.) If the mobile node can communicate only to one MAP (hard handover) then the address of the new MAP has to be sent to the old MAP just before the handover takes place but if the mobile is capable of communicating to more than one MAP simultaneously (soft handover) then the address can be sent any time during the handover. If the mobile suddenly loses the connection to the actual MAP then after finding new MAP it makes a normal handover in order to establish path to itself. The tracking handover in this case does not put any signalling load on the network except the load of the air (access) interface. For the cost functions see Table 2. The mobility management named Location Tracking (LTRACK) [8], introduced in 2003, belongs to this type of mobility management.

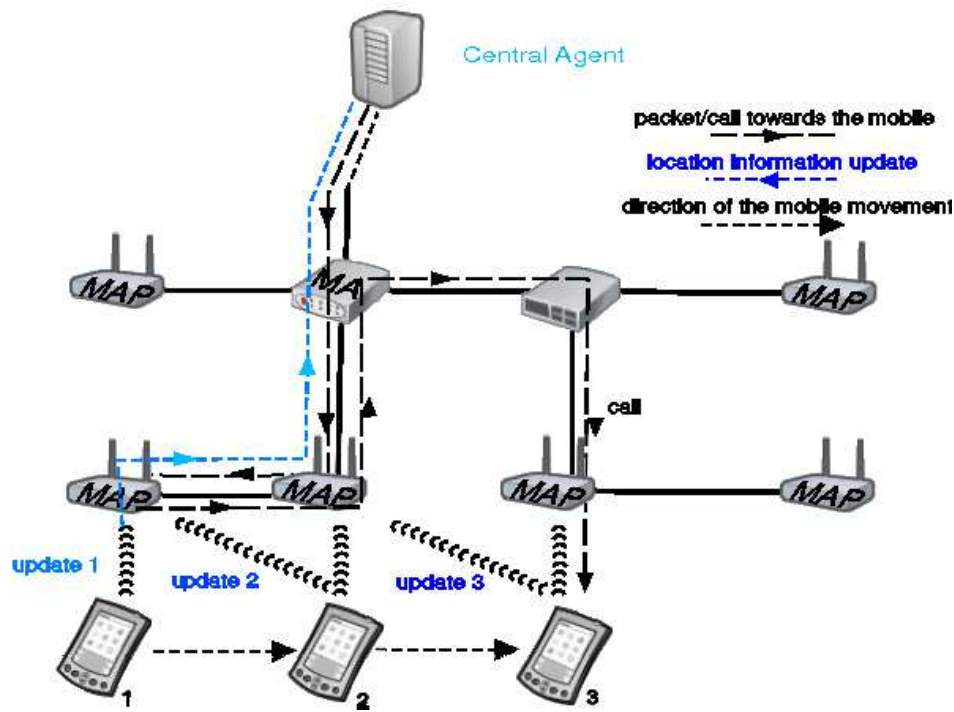

Fig. 4. Basic operation of Wireless tracking approach

### 4.3.2 Wired Tracking

Wired tracking differs from wireless one in the method of the tracking handover. In this case the mobile sends the address of the new MAP node to the old MAP node through the wired network like (HAWAII) [12] as it can be seen (Figure 5.) This handover puts some signalling load on the network, but it is not significant, and less than a signalling load in case of normal handover. The advantage of this method is that it saves the air interface resources. Partly similar cost functions to the wireless tracking can be derived and shown in Table 2. The well-known Handoff-Aware Wireless Access Internet Infrastructure (HAWAII) [12] is classable to the wired tracking algorithms.

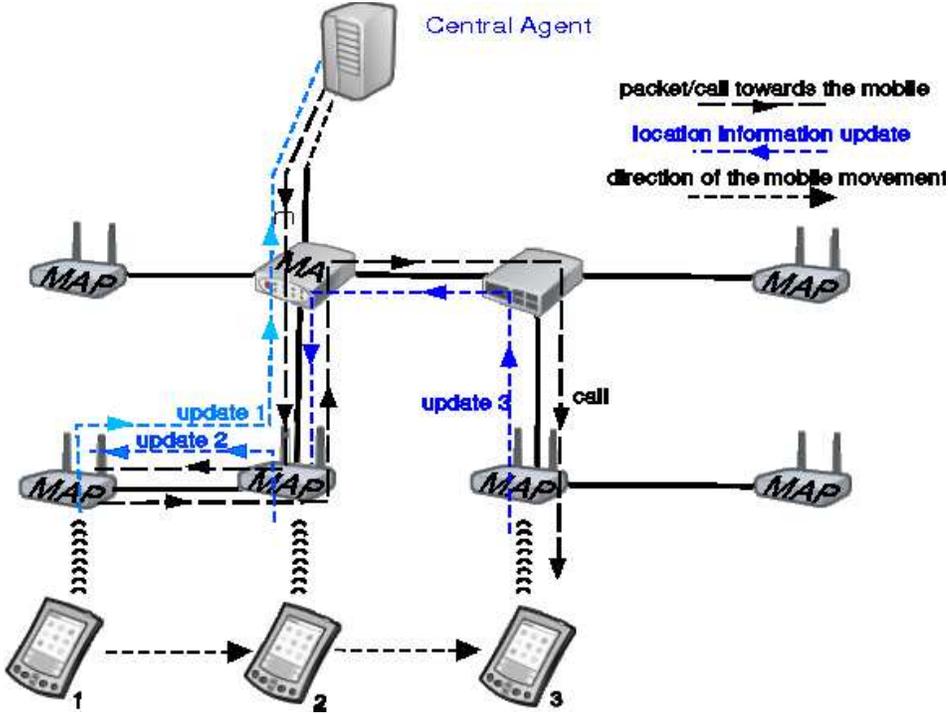

Fig. 5. Basic operation of Wired tracking approach

*4.3.3 Computing Parameters of Tracking Algorithms*

As we have seen when a normal handover occurs in tracking solutions, there is a signalling message sent to the Central Agent. It is obvious that a hierarchical mobility layer structure could be used in the way that it is used in every handover of the Hierarchical approaches. This is the reason why we used $g_H$ in the cost functions of tracking solutions. This makes the signalling cost of this solution lower than or equal to the cost of Hierarchical protocols. The processing cost could be less optimal because of the several decapsulation cost of the packet. To compute the number of optimal tracking handovers $H$ between normal handovers we adopt achievements from [8].

$$N = f(N) : min(C_{WLESSTRACKING}(H)) \qquad (6)$$

For the basic model the cost functions of wireless tracking can be treated as a continuous one and can be derived. This provides a fast and easy solution for computing the optimal value of $H$ with taking *looping* into account as well. If we extend our model with the effect of loop removal, it is clear that the cost function for tracking approaches remains the same, but values of the state probabilities $(P_H, P_0)$, and the expected point of return $(M[hr])$ will be different.

*4.4 Cellular-like Solutions*

To mobility problem there are cellular-like solutions as well. One well-known example is Cellular IP (CIP) [2]. The idea behind this kind of approach comes from the GSM protocol. Basically this kind of protocols can be used in lower level of the hierarchical network.

In a cellular systems the nodes do not know the topology of the network and the exact location of the MN. Packets to MN are routed by hop-by-hop manner, it is little the same as in tracking approaches. However, in most cellular solutions the routing is generally accomplished in layer 2, so the MAPs only have to know on which of its outgoing ports to forward packets. The solution builds strongly on the fact, that from the large number of mobile nodes only a small percentage are receiving data packets. For this reason we can define optimized -, well-defined areas, called paging area or location area, and it is enough to know in which paging the idle mobiles are moving. In this case the hop-by-hop manner routing leads the packet only to the domain border of the paging area. From this point of the network to the mobile the nodes in the paging area do not store any information about the idle mobiles, accordingly in case of packet addressed to an idle mobile the paging area is flooded with the packet by broadcast message. We have a general model for it, see Figure 6.

When an idle mobile realizes that the network searches for it then it changes its state to active, sends normal update to inform the paging area about its exact location, and receives the packets. After the normal update the broadcast sending becomes unnecessary. It is very important to perform this procedure quickly in order to the bearer network and the air interface could be spared the high cost of broadcast messages.

The advantage of this approach are the quick handover mechanism in lower layer and cheap passive connectivity as it can be seen through the cost functions on table 2 as well. The disadvantage is that the building of the network has to be careful and too many paging will cause an extreme increase in the costs. (One can see how the cost decreases with the growth of the mobility

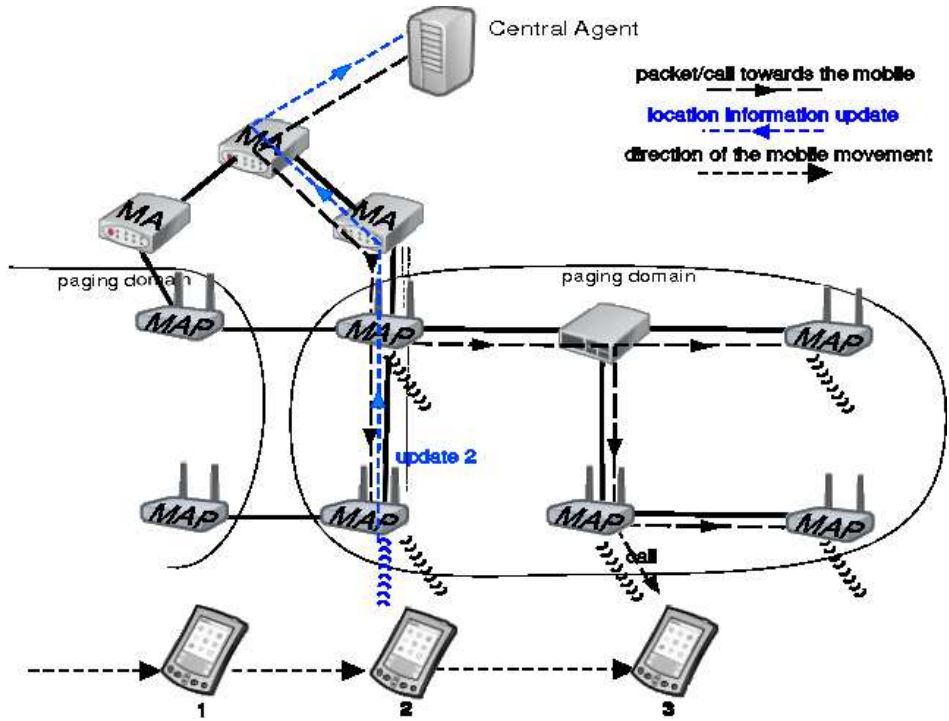

Fig. 6. Basic operation of Cellular approach

ratio $\rho$ in Section 5.)

### 4.4.1 Cellular-like Systems

The Cellular IP (CIP) is not the only one that uses the GSM like solution for mobility management. We will present another micro-mobility protocol based on the MANET [6] that uses the technique of wireless ad-hoc networks. However, the algorithm can be extended to macro-mobility level too and thus Hierarchical Paging [7] is introduced as an alternative Mobility Management solution.

The "MANET [6] in the paging areas" solutions introduced by us could be the best solution when we would like to save the infrastructure cost and the air interface using is cheaper. In this management system it is assumed that all MN could be reached via other MNs. Paging areas are defined like in other cell-like solutions, but only one MAP exists in one page, through this the packets are routed using an optimal MANET algorithm. Advantage of this solution also is that signaling cost can be saved with correct MANET protocol in a page. However, in the suboptimal case some mobiles could not be reached, and aggregate air interface cost can be high.

The main idea behind the Hierarchical Paging [7] is that not only the lower layer network is flooded with the packet but broadcast message is used to find the paging controller MA in the higher layer as well. With this functionality signalling cost could be saved because update messages are not sent to HA, but only to the MA which controls the page. But in case of calling the multilevel flooding causes high network load.

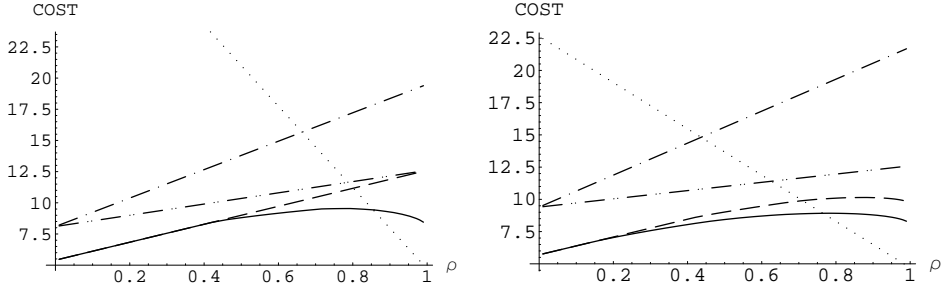

Fig. 7. One can se the summed cost functions of centralized-like (*one-dot-dash*), hierarchical-like (*two-dot-dash*), wireless (*dashed*) and wired (*solid*) tracking-like, cellular-like (*dotted*) approaches here with the vary of the mobility ratio: $\rho$. The two figures show the costs on different networks.

To discuss cellular approaches additional parameters have to be introduced to describe the paging area system. The ones we will use are: $P_C$ – in case of Cellular-like approaches the probability of entering to a new paging area; $P_M$ – in MANET-like solutions the request has to be sent via $P_M$ percent of mobile nodes at ad-hoc mobility level to be delivered it to the destination mobile node in a paging area. Two constants, related to the network topology are very important: $n_c$ – The average number of MAPs in a paging area; $n_d$ – The number of paging areas in the whole network.

Using the latter parameters all three type of mobility management can be modeled. All with similar cost functions as it can be seen in Table

## 5 Numerical Results

In this section we will present some general results on the models we have made and as a second part we try to answer some interesting questions for example: What if a cellular cell system was implemented on layer 3 level rather than layer 2? What if MIPv6 kind of messages and low level solutions were used for a Tracking-like approach? We will also show how our model can be used to make vertical handover decisions.

We do not attempt to give an exhausting numerical analysis with our method here since this paper focuses on the modeling framework itself. However, we give a very few examples for the type of investigations that could be performed using our model. The exact numerical values of the results are not important.

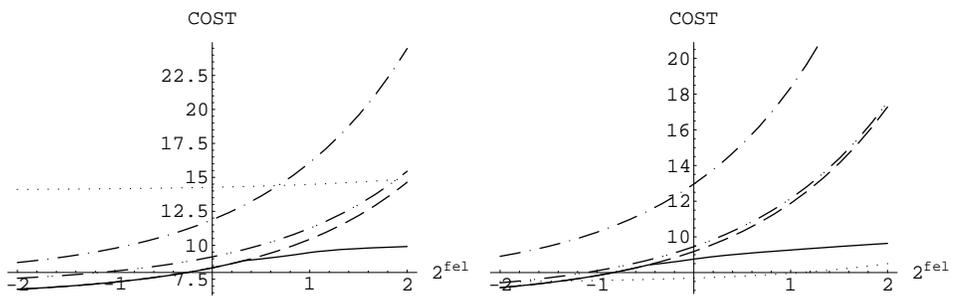

Fig. 8. The uplink/downlink vary dependency with the same notation at $\rho = 0.7$ and $\rho = 0.9$.

We focus on the behavior of mobility with the change of the parameters.

## 5.1 General Differences Between Management Approaches

On Fig. 7 one can see the difference between the approaches considering all the cost types (*signalling, processing, air*). It is clear that with the bigger frequency of handovers ($\rho$) the cost is bigger for the centralized-like, hierarchical-like and wired tracking-like approaches since each handover gives more signalling on the network. In the wireless tracking-like case if the number of handovers rise between the incoming calls it starts to save the costs of the rerouting of the packets. In the centralized-like ones, it is clear that the rarer there is an incoming call the less load the network has. The cost is obviously high in these case. The same case is printed on both figures, but the values of $g_T, g_C$ network parameters are significantly less than $g_H$ (more meshed network). One can see that the wired tracking-like solution is getting cheaper as well and begins to behave as its tracking-like pair.

On Fig. 8 the mobility ratio is fixed to ($\rho = 0.7, \rho = 0.9$) respectively. On the other hand the cost of a single upload ($c_u$) to a single download ($c_d$) is exponentially changing from the half to the twice on the horizontal axis. Most of the solutions are more expensive if the upload is higher but it can be seen that the wireless tracking cuts this cost as expected.

## 5.2 Technology Tendency via Cost Constants and Simulation

To answer questions like "What if a strategy like a cellular approach was implemented on a different technology level, we should give clear values of the *cost constants*. To do this we build up a simulation environment. OMNet++ [18] was used, which is a public-source, component-based, simulation environment with strong GUI support and an embeddable kernel. We have realized four IP layer mobility management, Mobile IPv4, Mobile IPv6 and our previous proposal, LTRACK on IPv4 and IPv6. We expect different technology constants for the two types of Internet Protocol however it is also unavoidable that two different kind of realized protocols differ in the constant values. During the simulation develop we followed strictly the protocols description and the result is presented in Table 3.

Shortly discussing the results we would like to point out an example. Taking a look at Table 3 it can be seen that in the case of MIPv6 the encapsulation cost ($c_{ec}$) is lower because there is no "triangle communication" instead it uses the mechanism of constantly updated "binding caches" while the registration cost ($c_d$) is higher because the IPv6 protocol overhead itself is higher. These facts intuitively verify our simulations. To prove the applicability of the simulation to our model framework we ran it on different network topologies and mobile movement setups and derived the expected value of each. With long run simulations the distribution concentrated near to the mean with low variation what proves the fact that these technology related constants are network and algorithm independent. (The network dependent cost constants for example the bandwidth measures are given in the weighted adjacency matrix $A$.)

Table 3
The cost constants of different technology levels from the simulation.

| Cost type | MIPv4 | MIPv6 | LTRACKv4 | LTRACKv6 |
|---|---|---|---|---|
| $c_u$ | 499 | 777 | 481 | 695 |
| $c_d$ | 1825 | 1821 | 1814 | 1843 |
| $c_r$ | 187 | 1068 | 711 | 821 |
| $c_f$ | 652 | 685 | 636 | 690 |
| $c_m$ | 543 | 1239 | 538 | 541 |
| $c_{ec}$ | 240 | 234 | 237 | 239 |
| $c_{rc}$ | 0 | 0 | 235 | 238 |
| $c_{dc}$ | 240 | 234 | 240 | 240 |
| $c_{au}$ | 136 | 181 | 138 | 165 |
| $c_{ad}$ | 390 | 396 | 368 | 397 |

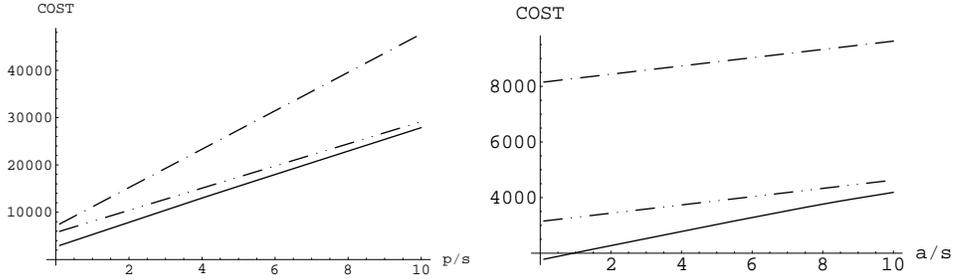

Fig. 9. The left Figure represents the cost of protocols in the ratio of "processing" and "signalling" costs while the right does the same in the ratio of "accessing" and "signalling" again.

In addition to the table above one is supposed to weight each type ("processing", "signalling", "accessing") of cost with different values since the cost of bandwidth is not in the same measure as the cost of processor load for example. On Figure 9 it can be followed for each how the cost of the mobility changes with the ratio of "processing" and "signalling" then the ratio of "accessing" and "signalling" cost respectively. If we know that in the future the mobile equipments and accessing technologies will become relatively cheap to the signaling resources then Figure 9 tells us that the ... protocol saves relatively the most with this tendency (of course on the given network and with the given implementation with the given mobility parameters, etc.).

## 5.3 How to Use the Model for Vertical Handover Decisions

We will present how to use our mobility management framework to make vertical handover decisions. Consider a vertical handover system of two protocols on the same region. The weights on network edges will denote the relative cost of using the given protocol divided by the QoS it provides. As an example, let network "A" be a locally maintained university WLAN network that can be used almost for free but provides bad QoS and network "B" be a GSM network covering the same area which provides better QoS but expensive. The ratio of the relative cost in money and some QoS measure depends on the user (or installed service). The network matrix than splits to two parts. One is for network "A" and the link weights (that basically shows relations between all the links) are multiplied by the above ratio expressing the user (anti-)preference. If the value ratio value is higher then the user is less willing to use the network since it has the more basic cost.

To grab the meaning of the vertical handover decision we have to manipulate the MN movement modeling part, the handover frequency matrix. If the network "A" is too expensive, then the total handovers from the network "B" are lower while backward handovers became higher expressing that the user is more willing to stay at the preferred part. Of course topologically the mobile nodes have to move within a given network too if they are making handovers.

In Figure 10 we depict an example how the vertical handover decision can be modeled within our framework. The two networks can be found on the top and the bottom. The nodes can move within the access points (MAP) of a network as a part of the normal operation. Note that the two Access network do not need to cover the same geographical area instead where vertical handover is possible we put a link into the Handover Frequency graph with $\nu$ and $1/\nu$ for each direction.

As one can realize we used a stochastic method to make vertical handover decisions. It is in the scope of our research to compare it to others but for now we try to emphasize why a stochastic method should be applied through an example. Suppose that the cost of a required QoS on the nodes are as it is depicted on the graph on Figure 10. The total cost of QoS in access network "A" is higher than in "B". On the other hand if the user mobility is higher than while it receives less traffic it is much more worth to stay at network "A". Since the mobility is a Poisson process, and the threshold depends on it (not just on its mean), it will vary according to a random variable too.

We used *Mathematica* for the implementation of the cost functions and the modeling and its capability for symbolic computations helps us to easily manipulate those rows and columns in the handover frequency matrix which describe the above motions and relations. In Figure 11 one can see how the final cost of mobility changes with the vertical handover willing rate ($\nu$) from network "A" to network "B". The point where $\nu = 1$ shows the cost for the situation when it is irrelevant for the MN which network it uses. One can see that it is not (necessary) the point with the optimal mobility cost $\nu^*$. In this current example the mobility cost is interpreted as a utility function for the user. The user is willing to optimize its cost and QoS ratio so will always set $\nu = \nu^*$ when making a handover and chooses network "A" with probability $P[\text{``Choose Network ``A''.''}] = (\nu\tau) \exp(\nu\tau)$, since $\nu$ is the intensity of the Poisson process of making a handover.

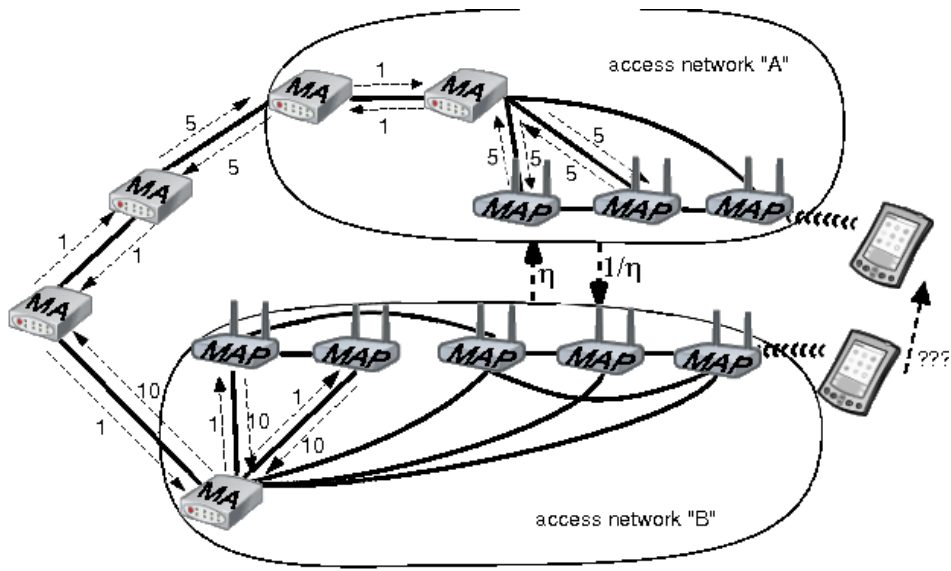

Fig. 10. Vertical handover modelling example

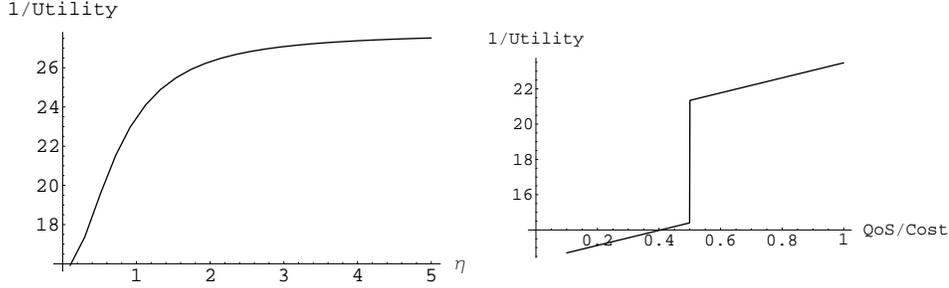

Fig. 11. On the left hand side one can see the 1/Utility (Cost) variation for the user as a function of $\nu$, handover probability to network "A" while on the left hand side figure one can see the 1/Utility (Cost) when the need for QoS/Cost varies.

The interpretation is the following. Let event $E_1$, $E_2$ be "Vertical handover to Network "A" before making a handover.", "Not receiving a call." respectively thus $P[E_1|E_2] = \frac{\nu}{\nu + \sum_{\forall i} w_i}$ were $w_i$ is the handover intensity to a given direction. Another interpretation can be done with the events $E_3$ and $E_4$ meaning "Handover to Network "A" before receiving the next call." and "Make no handover within Network "B"." respectively thus $P[E_3|E_4] = \frac{\nu}{\nu+\mu}$. The latest interpretations are important for us to understand the parameter $\nu$. It effectively tells us the intensity of moving from Network "A" to Network "B". If there was such a handover, than the backward handover is expected, with intensity of $\frac{1}{\nu}$.

Of course as the quality in a network falls the user is more likely to switch to the other. In Figure 11 we depict the change of Utility in the function of the ratio of cost and quality separately while in Figure 12 we depict them for Mobile IP and Hierarchical Mobile IP according to both two parameters. In most of the QoS/Cost parameter cases one might find the optimal $\nu^*$ rate going to infinity which means that it is worth to switch network immediately while it is not worth to switch back. Proofs of limits of $\nu$ in infinity is a subject of another work.

To understand the above calculations we would like to point out a few interesting questions.

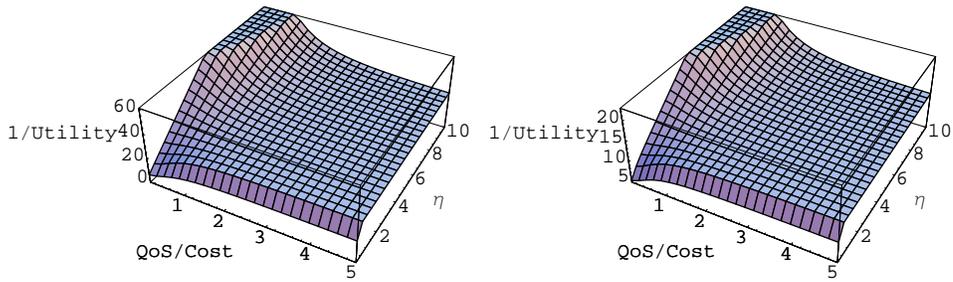

Fig. 12. On the left hand side figure one can see the 1/Utility (Cost) variation for the user using MIP protocol while on the right hand side for HMIP protocol. Both are depicted as a function of $\nu$, handover probability to network "A" and QoS/Cost. (The shape of curves are similar but the values are not.)

The first to answer is: "How does the single network matrix and the derived parameter $m, g_H$, etc. differentiate between the two networks?" The answer is it does via the stationary distribution of the Mobile Nodes. If the $\nu$ is high than handover rate from Network "A" to "B" is much higher than backwards. This means that the MNs spend more time at the Access Points of Network "B". When we calculate $m$ we multiply the shortest paths in the network graph with this probability distributions thus paths with higher Utility (the paths of Network "B") will have bigger weights that says we use them more often.

One might believe that the intensity of handovers to the network with the bigger Utility is higher than backwards but "Why do we have handovers to the network with less Utility at all?". The answer for this question is obvious: because we have used a Stochastic model. We have never stated that Network "A" is better than Network "B" in terms of QoS to Cost ratio, but we have stated that it is better with respect to the stochastic model we used (stochastic processes for both call arrival and handover rate). It is not a part of our current work to give analytical proof of the fact that our stochastic model is better for vertical handover decisions than existing deterministic ones.

## 6 Conclusions and Future Work

In this paper we grabbed numerous significant parameters of mobility and proposed a framework to model the mobile node behavior and the network independently of the very technology used. As an example we modeled some general management strategies and obtained technology constants with simulations to give some analytical results. We showed how each modeled protocol responds to tendencies in technology developments and trends and proposed a method to deal with vertical handover decisions using the framework. With our work it could be shown which mobility management gives the best solution in different given network scenarios and which aspect of resources could be a bottleneck in each case. This can help operators to tune the parameters of their systems, plan their networks and researchers to propose mobility management solutions of low cost.

**Acknowledgements**

Special thanks to the support from the High Speed Network Laboratory: *http://www.tmit.bme.hu/labgroup/hsn!eng.*

# References

[1] Abondo, C., & Pierre, S., Dynamic Location And Forwarding Pointers For Mobility Management, 2005, Mobile Information Systems, IOS Press, 3-24.

[2] Campbell, A. T., Gomez, J., & Valkó, A. G., An Overview of Cellural IP. 1999, IEEE., 29-34.

[3] Castelluccia, C., A Hierarchical Mobile IP Proposal, 1998, Inria Technical Report.

[4] Fang, Y., & Lin, Y., Portable Movement Modelling for PCS Networks, 2000, IEEE Transactions on Vehicular Technology, 1356-1362.

[5] Jokela, P., Nikander, P., Melen, J., Ylitalo, J., & Wall, J., Host Identity Protocol: Achieving IPv4 - IPv6 handovers without tunneling, 2003, Proceedings of Evolute workshop, "Beyond 3G Evolution of Systems and Services", University of Surrey, Guildford, UK.

[6] Ashwini K. Pandey, Hiroshi Fujinoki, Study of MANET routing protocols by GloMoSim simulator International 2005, Journal of Network Management.

[7] Szalay, M., Imre, S., Hierarchical Paging - A novel location management algorithm, 2006, ICLAN'2006 International Conference on Late Advances in Networks, Paris, France.

[8] Kovács, B., Szalay, M., & Imre, S., Modelling and Quantitative Analysis of LTRACK - A Novel Mobility Management Algorithm, 2006, Mobile Information Systems, Issue: Volume 2, Number 1/2006, 21 - 50.

[9] Ma, W., & Fang, Y., Dynamic Hierarchical Mobility Management Strategy for Mobile IP Networks, 2004, IEEE Journal of Selected Areas In Communications.

[10] Perkins, C. E., Mobile IP, 1997, IEEE Communications Magazine.

[11] Escalle, P. G., Giner, V. C., & Oltra, J. M., Reducing Location Update and Paging Costs in a PCS Network, 2002, IEEE Transactions on Wireless Communications, 200-209.

[12] Ramjee, R., La Porta, T., Thuel, S., Varadhan, K., & Salgarelli, L., A Hierarchical Mobile IP Proposal, 1998, Inria Technical Report.

[13] S. Das, A., Misra, P. Agraval & S. K. Das, TeleMIP: Telecommunications-Enhanced Mobile IP Architecture for Fast Intradomain Mobility, 2000, IEEE Personal Communications, 50-58.

[14] Vilmos Simon, Sndor Imre, Location Area Design Algorithms for Minimizing Signalling Costs in Mobile Networks, International Journal of Business Data, 2007, Communications and Networking.

[15] Szalay, M., Imre, S., *Hierarchical Paging- Efficient Location Management*, 2007, Fourth European Conference on Universal Multiservice Networks (ECUMN'07), 301-310.

[16] Vilmos Simon, Sndor Imre, A Paging Cost Constrained Location Area Planning in Next Generation Mobile Networks, 2006, The 4th International Conference on Advances in Mobile Multimedia- MoMM 2006.

[17] Wolfram Research Inc.: *Mathematica*. http://www.wolfram.com/.

[18] http://www.omnetpp.org/